%% file: 0_main.tex
\documentclass[sigconf]{acmart}

\AtBeginDocument{%
  \providecommand\BibTeX{{%
    \normalfont B\kern-0.5em{\scshape i\kern-0.25em b}\kern-0.8em\TeX}}}

\usepackage{booktabs}
\usepackage{array}
\usepackage{balance} 
\usepackage{multirow}
\usepackage[normalem]{ulem}
\usepackage{color}
\definecolor{lightgray}{RGB}{215,215,215}
\usepackage{colortbl}  
\usepackage{xcolor}
\useunder{\uline}{\ul}{}
\usepackage{subfigure}
\usepackage{algorithm}  
\usepackage{algorithmicx}  
\usepackage[noend]{algpseudocode}  

\usepackage{amsmath}  
\usepackage{enumitem}
\usepackage{tabularx}
\usepackage[utf8]{inputenc}
\usepackage[english]{babel}
\usepackage{amsthm}
\usepackage{bm}
\newcommand{\ie}{\emph{i.e., }}
\newcommand{\eg}{\emph{e.g., }}

\newcommand{\cf}{\emph{cf. }}

\newlength\myindent
\floatname{algorithm}{Algorithm}

\copyrightyear{2024}
\acmYear{2024}
\setcopyright{acmlicensed}\acmConference[KDD '24]{Proceedings of the 30th ACM SIGKDD Conference on Knowledge Discovery and Data Mining}{August 25--29, 2024}{Barcelona, Spain}
\acmBooktitle{Proceedings of the 30th ACM SIGKDD Conference on Knowledge Discovery and Data Mining (KDD '24), August 25--29, 2024, Barcelona, Spain}
\acmDOI{10.1145/3637528.3671884}
\acmISBN{979-8-4007-0490-1/24/08}

\begin{document}


\title{Bridging Items and Language: A Transition Paradigm for Large Language Model-Based Recommendation}

\author{Xinyu Lin}
\email{xylin1028@gmail.com}
\affiliation{
\institution{National University of Singapore}
\city{}
\country{Singapore}
}
\author{Wenjie Wang$^*$}
\email{wenjiewang96@gmail.com}
\affiliation{
\institution{National University of Singapore}
\country{Singapore}
}
\author{Yongqi Li}
\email{liyongqi0@gmail.com}
\affiliation{
\institution{The Hong Kong Polytechnic University}
\city{Hong Kong SAR}
\country{China}
}

\author{Fuli Feng}
\email{fulifeng93@gmail.com}
\authornote{Corresponding author. This research is supported by A*STAR, CISCO Systems (USA) Pte. Ltd and National University of Singapore under its Cisco-NUS Accelerated Digital Economy Corporate Laboratory (Award I21001E0002), and the National Natural Science Foundation of China (62272437).}
\affiliation{
\institution{University of Science and Technology of China}
\city{Hefei}
\country{China}
}
\author{See-Kiong Ng}
\email{seekiong@nus.edu.sg}
\affiliation{
\institution{National University of Singapore}
\city{}
\country{Singapore}
}
\author{Tat-Seng Chua}
\email{dcscts@nus.edu.sg}
\affiliation{
\institution{National University of Singapore}
\city{}
\country{Singapore}
}

\renewcommand{\shortauthors}{Xinyu Lin et al.}
\begin{abstract}
Harnessing Large Language Models (LLMs) for recommendation is rapidly emerging, which relies on two fundamental steps to bridge the recommendation item space and the language space: 1) \textbf{item indexing} utilizes identifiers to represent items in the language space, and 2) \textbf{generation grounding} associates LLMs' generated token sequences to in-corpus items. However, previous methods exhibit inherent limitations in the two steps. Existing ID-based identifiers (\eg numeric IDs) and description-based identifiers (\eg titles) either lose semantics or lack adequate distinctiveness. Moreover, prior generation grounding methods might generate invalid identifiers, thus misaligning with in-corpus items.

To address these issues, we propose a novel \textbf{Trans}ition paradigm for LLM-based \textbf{Rec}ommender (named TransRec) to bridge items and language. Specifically, TransRec presents multi-facet identifiers, which simultaneously incorporate ID, title, and attribute for item indexing to pursue both distinctiveness and semantics. Additionally, we introduce a specialized data structure for TransRec to ensure generating valid identifiers only and utilize substring indexing to encourage LLMs to generate from any position of identifiers. Lastly, TransRec presents an aggregated grounding module to leverage generated multi-facet identifiers to rank in-corpus items efficiently. We instantiate TransRec on two backbone models, BART-large and LLaMA-7B. 
Extensive results on three real-world datasets under diverse settings validate the superiority of TransRec. 

\vspace{-0.2cm}
\end{abstract}

\maketitle

\input{1_intro}
\input{2_0_preliminaries}
\input{2_method}
\input{3_exp}

\input{4_related_work}
\input{5_conclusion}
\clearpage

{
\tiny
\bibliographystyle{ACM-Reference-Format}
\balance
\bibliography{bibfile}
}

\appendix
\input{6_SI}
\clearpage

\end{document}

%% file: 1_intro.tex
\section{Introduction}\label{sec:introduction}

\begin{figure}[t]
\setlength{\abovecaptionskip}{0cm}
\setlength{\belowcaptionskip}{-0.4cm}
\centering
\includegraphics[scale=1.2]{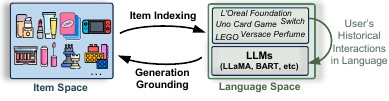}
\caption{{Illustration of the two pivotal steps for LLM-based recommenders: item indexing and generation grounding.}}
\label{fig:intro_2stage}
\end{figure}

Large Language Models (LLMs) have achieved remarkable success across diverse domains~\cite{hong2022bros, berrios2023towards, liang2023code} due to their emergent competencies, including possessing rich knowledge~\cite{guu2020retrieval}, instruction following~\cite{ouyang2022training}, and in-context learning~\cite{kojima2022large}. 
Recently, there has been a notable surge in exploring the benefits of adapting LLMs to recommendations. 
In particular, LLMs have showcased the potential in discerning nuanced item semantics~\cite{zhang2021unbert,wu2022mm}, understanding multiple user interests~\cite{deng2022toward,zhang2023recommendation}, and generalizing to cold-start item recommendations~\cite{bao2023tallrec,chen2023palr,gao2023chat}. 
In light of these, 
the prospect of harnessing LLMs as recommender systems, \ie LLM-based recommenders, emerges as a particularly promising avenue for further exploration.

Given that recommender systems and LLMs work in the item space and the language space, respectively, the key to building LLM-based recommenders lies in bridging the item space and the language space (refer to Figure~\ref{fig:intro_2stage}). 
Here, the item space includes all the existing items on the recommender platform. 
To bridge the two spaces, it involves two essential steps: 
item indexing and generation grounding. 
The item indexing step assigns each item with a unique identifier (\eg item title or numeric ID) in natural language, and subsequently the user's historical interactions are converted into a sequence of identifiers. 
To yield recommendations, given the identifier sequence of the user's historical interactions, the generation grounding step 
utilizes LLMs to generate a token sequence in the language space and inversely ground the token sequence to the items in the item space. 
However, existing work has intrinsic limitations in both steps.

\textbf{Item indexing.} 
Previous studies can be categorized into ID-based identifiers~\cite{geng2022recommendation,hua2023index} and description-based identifiers~\cite{bao2023tallrec,cui2022m6}.
\begin{itemize}[leftmargin=*]
    \item ID-based identifiers utilize numeric IDs (\eg ``15308'') to represent items, effectively capturing the uniqueness of items~\cite{geng2022recommendation,hua2023index}. 
    Nevertheless, IDs lack explicit semantics\footnote{Item indexing methods that aim to integrate semantics into numeric IDs are compared and discussed in Section~\ref{sec:overall_performance} and Section~\ref{sec:related_work}.} and hinder the knowledge generalization of LLMs. 
    Worse still, LLMs require sufficient interactions to fine-tune each ID identifier, decreasing the generalization ability to large-scale and cold-start recommendations. 
    \item Description-based identifiers adopt semantic descriptions (\eg titles and attributes) to index items~\cite{bao2023tallrec,cui2022m6}. 
However, item descriptions lack adequate distinctiveness due to the existence of common words (\eg ``Will'' and ``Be'' in movie titles). Moreover, item descriptions might not consistently align with user-item interactions: two items with similar descriptions may not have similar interactions. This divergence is a possible reason for the wide usage of IDs in existing feature-based recommendations. 
\end{itemize}

\textbf{Generation grounding.} 
In previous work, LLMs autoregressively generate a sequence of tokens via beam search, and then ground the token sequence to the identifiers of items via exact matching~\cite{geng2022recommendation}. 
Nevertheless, unconstrained generation over the whole vocabulary of LLMs may yield invalid item identifiers, leading to out-of-corpus recommendations~\cite{liu2023chatgpt}. 
Thus, 
additional matching strategies (\eg L2 distance matching~\cite{bao2023bi}) are necessary to ground the out-of-corpus identifiers to existing identifiers, which however is computationally expensive (\cf Section~\ref{sec:ablation}). 
Drawing upon the above insights, we establish some key objectives for the two steps. 
For \textbf{item indexing}, we posit that item identifiers should at least satisfy two criteria: 
1) \textit{distinctiveness} that ensures the items are distinguishable from each other; and 2) \textit{semantics} that guarantees the full utilization of the rich knowledge in LLMs, enhancing the generalization abilities of LLM-based recommenders. 
For \textbf{generation grounding}, we consider constrained generation to ensure generating valid identifiers without additional matching. 
Nonetheless, a crucial challenge is that constrained generation heavily relies on the generation quality of the initial tokens during beam search. 
Because constrained generation strictly starts from the first token of valid identifiers~\cite{chu2023leveraging}, LLMs cannot generate the ideal item if the generated first token is incorrect.
As such, we 
consider 
\textit{position-free constrained generation} to allow LLMs to generate from any position in the valid identifiers.  




\begin{figure}[t]
\setlength{\abovecaptionskip}{0cm}
\setlength{\belowcaptionskip}{-0.3cm}
\centering
\includegraphics[scale=1.3]{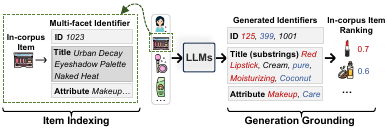}
\caption{{Overview of TransRec. Item indexing assigns each item a multi-facet identifier. For generation grounding, TransRec generates a set of identifiers in each facet and then grounds them to in-corpus items for ranking.}}
\label{fig:intro_ourmethod}
\end{figure}
To improve item indexing and generation grounding as the bridge between the item and language spaces, we propose a novel transition paradigm for LLM-based recommenders (shorted as TransRec). 
Specifically, 1) as depicted in Figure~\ref{fig:intro_ourmethod}(a), TransRec indexes items with multi-facet identifiers, which simultaneously considers item IDs, titles, and attributes (\eg category) as the identifiers in separate facets to pursue both distinctiveness and semantics. 
And 2) for generation grounding, 
to achieve position-free constrained generation, 
we introduce FM-index, a data structure that supports LLMs to generate any segment of valid identifiers. 
Besides, to further enhance the position-free generation ability of LLMs, 
we propose also using substrings of identifiers to index items (\eg ``Lipstick-Red'' in ``Everyday Elegance Lipstick-Red'') for instruction tuning. 
Lastly, LLMs will generate valid identifiers in each facet as in Figure~\ref{fig:intro_ourmethod}(b), and we present an aggregated grounding module to leverage generated identifiers to rank in-corpus items. 
To validate the effectiveness of TransRec, we conduct extensive experiments on three real-world datasets under diverse settings, including full training and few-shot training with warm- and col-start testings. Empirical results on two backbone models BART-large~\cite{lewis2019bart} and LLaMA-7B~\cite{touvron2023llama} reveal the superiority of TransRec over traditional models and LLM-based models. 
The code and data are at \url{https://github.com/Linxyhaha/TransRec/}. 

In summary, this work offers several significant contributions:
\begin{itemize}[leftmargin=*]
    \item We highlight existing problems and the key objectives in the two fundamental steps of bridging the item and language spaces in LLM-based recommenders. 
    \item We propose a new TransRec paradigm with multi-facet identifiers and position-free constrained generation, seamlessly bridging items and language for LLM-based recommendations. 
    \item We conduct extensive experiments under various recommendation settings, demonstrating the effectiveness of TransRec. 
\end{itemize}

%% file: 2_0_preliminaries.tex
\section{Preliminary}\label{sec:preliminaries}
This section first introduces the process of establishing LLM-based recommenders, including instruction tuning and generation grounding. 
And then, we highlight the two key steps to bridge between the item space and the language space, and reveal the vulnerabilities of existing work. 

\vspace{3pt}
\noindent$\bullet\quad$\textbf{Task formulation.}
Let $\mathcal{U}$ and $\mathcal{I}$ denote the sets of users and items, respectively. 
We represent the historical interaction sequence of a user $u$ by $\mathcal{S}_u=[i_u^1, i_u^2, \dots, i_u^{L}]$ in a chronological order, where $u\in \mathcal{U}$, $i_u\in \mathcal{I}$, and $L=|\mathcal{S}_u|$. 
Given a user's historical interactions, the sequential recommendation task\footnote{We mainly focus on sequential recommendation as it considers the crucial temporal aspect in real-world scenarios, showing notable practical significance~\cite{xie2021adversarial,xie2022decoupled,tian2022multi}.} is to predict this user's next liked item $i_u^{L+1}\in \mathcal{I}$. 


To leverage LLMs for recommendation, 
existing work mainly resorts to instruction tuning, to narrow the intrinsic discrepancy between the LLMs' pre-training data and the recommendation data. 
After instruction tuning, LLMs are instructed to generate recommended items based on the user's historical interactions. 
In the following, we introduce the preliminary knowledge of instruction tuning and generation grounding. 

\begin{figure}[t]
\vspace{-0.3cm}
\setlength{\abovecaptionskip}{-0.15cm}
\setlength{\belowcaptionskip}{-0.1cm}
\centering
\includegraphics[scale=1.35]{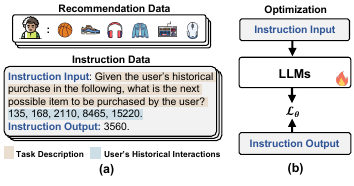}
\caption{{Illustration of instruction tuning of LLMs. (a) depicts the conversion from recommendation data to instruction data; (b) presents the optimization of LLMs based on the instruction data.}}
\vspace{-0.3cm}
\label{fig:instruction_tuning}
\end{figure}

\subsection{Instruction Tuning}

Instruction tuning involves three phases: item indexing, data reconstruction, and LLM optimization.  

\vspace{2pt}
\noindent$\bullet\ $\textbf{Item indexing.} 
For each item $i\in\mathcal{I}$ in recommendation data, item indexing assigns an identifier $\Tilde{i}\in\Tilde{\mathcal{I}}$ in natural language, where $\Tilde{\mathcal{I}}$ is the identifier corpus. 
To achieve item indexing, exiting work can be categorized into ID-based identifiers (\eg numeric IDs~\cite{geng2022recommendation,hua2023index}), and description-based identifiers (\eg titles~\cite{bao2023bi} and attributes~\cite{cui2022m6}). 

\vspace{2pt}
\noindent$\bullet\ $\textbf{Data reconstruction}. Based on the item identifiers, each user's interaction sequence $\mathcal{S}_u=[i_u^1, i_u^2, \dots, i_u^{L}]$ could be converted to a sequence of identifiers $\Tilde{\mathcal{S}_u}=[\Tilde{i}_u^1, \Tilde{i}_u^2, \dots, \Tilde{i}_u^{L}]$. 
Thereafter, instruction data $\mathcal{D}_{instruct}=\{(x,y)\}$ are constructed for instruction tuning, where $x$ and $y$ 
denote the instruction input and output, respectively. 
Specifically, as shown in Figure~\ref{fig:instruction_tuning}(a), 
the instruction input contains the \textit{task description} illustrating the recommendation task; and the \textit{user's historical interactions} $[\Tilde{i}_u^1, \Tilde{i}_u^2, \dots, \Tilde{i}_u^{L-1}]$ in natural language. 
The instruction output is usually set to the identifier of the next-interacted item, \ie $y=\Tilde{i}_u^{L}$. 

\vspace{2pt}
\noindent$\bullet\ $\textbf{LLM optimization}. 
Given the instruction data $\mathcal{D}_{{instruct}}$, the learnable parameters ($\theta\in\Theta$) of an LLM can be optimized by minimizing the negative log-likelihood of instruction output $y$ conditioned on input $x$: 
\begin{equation}\small\label{eq:pre_finetuning_objective}
    \mathop{\min}_{\theta\in\Theta} \{\mathcal{L}_{\theta}=-\sum_{t=1}^{|y|}\log P_{\theta}(y_t|y_{<t},x)\}, 
\end{equation}
where $y_t$ is the $t$-th token of $y$, and $y_{<t}$ represents the token sequence  preceding $y_t$. 
\subsection{Generation Grounding} 
After the instruction tuning, we can effectively leverage LLMs to generate recommendations via the generation grounding step, \ie generating token sequences by LLMs and grounding them to the in-corpus items. 

\vspace{2pt}
\noindent$\bullet\ $\textbf{Generation.} 
Given an instruction input $x$, which contains the \textit{task description} and the \textit{user's historical interactions} $[\Tilde{i}_u^1, \Tilde{i}_u^2, \dots, \Tilde{i}_u^{L}]$, LLM-based recommender autoregressively generates a token sequence $\hat{y}$ step by step via beam search. 
Formally, when beam size $= 1$, at each time step $t$, we have 
\begin{equation}\small\label{eq:pre_autoregressive}
    \hat{y}_t = \mathop{\arg\max}_{v\in\mathcal{V}} P_{\theta}(v|\hat{y}_{<t},x), 
\end{equation}
where $\mathcal{V}$ is the token vocabulary of the LLM. The LLM-based recommender keeps generating until it meets stopping criteria (\eg $\hat{y}_t$ is the stop token ``EOS''). 

\vspace{2pt}
\noindent$\bullet\ $\textbf{Grounding.} 
The generated token sequence $\hat{y}$ in the language space is then grounded to a set of existing identifiers as recommendations: 
\begin{equation}\small
\label{eq:pre_grounding}\notag
    \{\Tilde{i}|\Tilde{i}\in\Tilde{\mathcal{I}}\} \leftarrow \text{Ground}(\hat{y}), 
\end{equation}
where $\text{Ground}(\cdot)$ is the grounding approach, such as exact matching~\cite{geng2022recommendation}, and distance-based matching~\cite{bao2023bi}. 

\vspace{2pt}
To sum up, bridging the item space and the language space for building LLM-based recommenders involves two fundamental steps: \textbf{\textit{item indexing}} and \textbf{\textit{generation grounding}}. 
{Upon the two steps, LLMs can follow the standard operations of data reconstruction and instruction tuning in the language space. }
However, previous work suffers from intrinsic limitations in the two steps.
For item indexing, 
existing ID-based identifiers and description-based identifiers either lose semantics or lack adequate distinctiveness, leading to the underutilization of rich knowledge in LLMs or losing salient features crucial to the recommendation. 
{Notably, little effort has been made to study or compare the ID- and description-based identifiers specifically for LLM-based recommendation (related work is discussed in Section~\ref{sec:related_work}), where the potential enhancement from the incorporation of both IDs and semantics specifically for LLM-based recommendation deserves more exploration.} 
For generation grounding, 
Eq. (\ref{eq:pre_autoregressive}) allows for generating any token from the LLMs' vocabulary at each step, potentially leading to out-of-corpus identifiers.
Additional matching approaches~\cite {bao2023bi} can mitigate the out-of-corpus issue, which however is time-consuming (\cf Section~\ref{sec:ablation}). 


%% file: 2_method.tex
\section{Method}\label{sec:method}
To strengthen LLM-based recommenders from the two crucial steps, we propose a novel transition paradigm TransRec, which involves multi-facet item indexing and generation grounding. 

\subsection{Multi-facet Item Indexing} 
To alleviate the intrinsic limitations of existing item indexing methods, we postulate two criteria for item identifiers: 1) distinctiveness to ensure the items are distinguishable from each other; and 2) semantics to make full utilization of rich knowledge in LLMs, enhancing the generalization abilities. 

\subsubsection{\textbf{Multi-facet Identifier}}\label{sec:multi-facet-identifier}

Well meeting the above two criteria, we propose multi-facet identifiers for the item indexing step. 
In particular, we simultaneously incorporate three facets to represent an item from different aspects: 

\noindent$\bullet\ $\textbf{Numeric ID} 
guarantees the distinctiveness among items. 
{We assign each item a unique random numeric ID, denoted by $P$ (\eg ``3471''). }
By tuning over user-item interactions described by unique IDs, LLMs are incentivized to align the numeric IDs with the user-item interactions, {which can capture crucial Collaborative Filtering (CF) knowledge}. 
In essence, items with similar interactions are endowed with similar ID representations. 

\noindent$\bullet\ $\textbf{Item title} ensures rich semantics that can capitalize the wealth of world knowledge in LLMs. 
An item title, denoted by $T$, \eg ``Rouge Coco Hydrating Creme Lipstick Chanel \#432'', typically contains a concise and descriptive name, conveying some general information about the item. 


\noindent$\bullet\ $\textbf{Item attribute} serves as a complementary facet to inject semantics, particularly in cases where item titles may be less informative or unavailable. 
For an item that has multiple attributes such as ``Makeup, Eyes, Multicolor, Gluten Free'', we denote each attribute as $a$ and the complete attributes as $A=[a_1, a_2, \dots, a_{n}]$. 

\vspace{2pt}
In summary, for each item $i$ in the recommendation data, we can obtain the multi-facet identifier $\Tilde{i}=\{P, T, A\}$. 
Based on the multi-facet identifiers, we then construct the instruction data in language space for the instruction tuning of LLMs.

\subsubsection{\textbf{Data Reconstruction}}\label{sec:multi-finetuning}

\begin{figure}[t]
\vspace{-0.5cm}
\setlength{\abovecaptionskip}{0.05cm}
\setlength{\belowcaptionskip}{-0.1cm}
\centering
\includegraphics[scale=0.38]{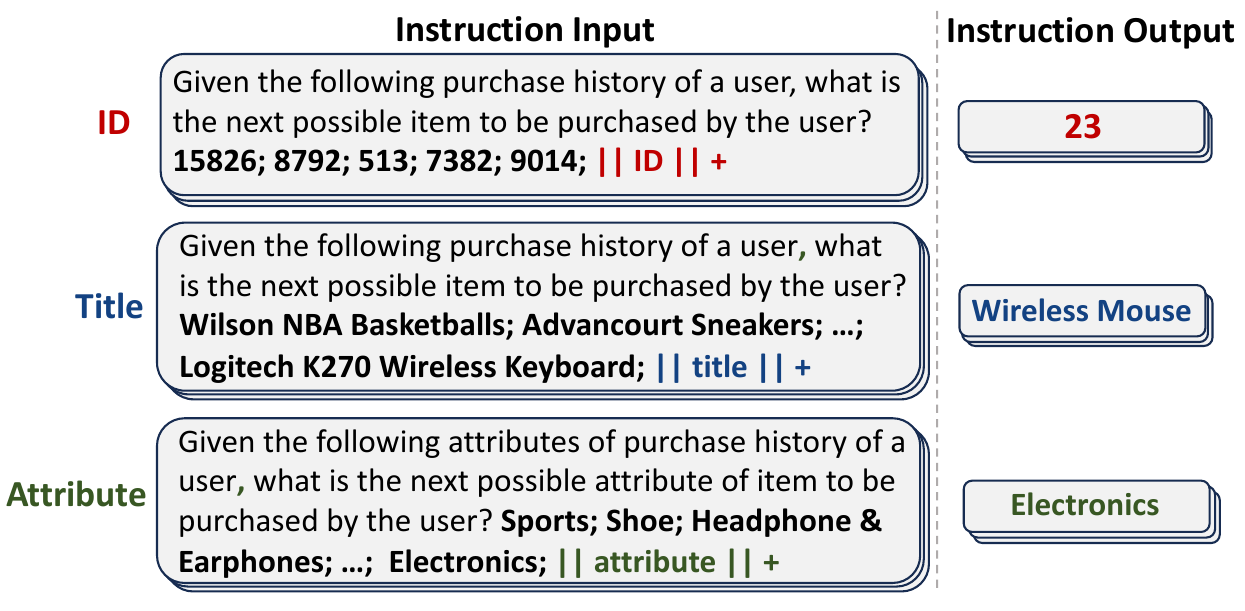}
\caption{{Illustration of the reconstructed data based on the multi-facet identifiers. The bold texts in black refer to the user's historical interactions.}}
\label{fig:reconstruced_data}
\end{figure}

As shown in Figure~\ref{fig:reconstruced_data}, we convert each user's interaction sequence $\mathcal{S}_u=[i_u^1, i_u^2, \dots, i_u^{L}]$ into instruction data in three facets, separately. 
For each facet, we construct instruction input and output based on the user's interaction sequence. 
We fix the templates of task descriptions\footnote{Full templates can be found at https://github.com/Linxyhaha/TransRec/.}, and mainly focus on the reconstruction of the user's historical interactions and the instruction output in the following. 


To form the user's historical interactions for the ID facet, 
we convert the first $L-1$ items in $\mathcal{S}_u$ to their numeric IDs, and then separate each item with a semicolon. A sequence of ID-facet identifiers is denoted as ``$P_1\text{; }P_2\text{; }\dots\text{; }P_{L-1}$''\footnote{For notation brevity, we omit the subscript $u$ representing the user.}. 
Likewise, we can obtain the user's historical interactions in the title and attribute facets, referred to as ``$T_1\text{; }T_2\text{; }\dots\text{; }T_{L-1}$'', and ``$A_1\text{; }A_2\text{; }\dots\text{; }A_{L-1}$'', respectively. 
As for the instruction output, for the ID facet, we use the numeric ID of the last item in the user's interaction sequence, \ie $P_L$. 
For the title facet, 
we utilize substrings $t$ with arbitrary length $l\in\{1,\dots,|T|\}$, sampled from the title $T$. 
This is to encourage LLMs to generate from any positions that are possibly relevant to the user's interests. 
Here, for each user's interaction sequence, we sample $K$ substrings of the last item's title and construct $K$ instruction input-output pairs. 
Lastly, for the attribute facet, each attribute $a\in A_L$ is independently used as one instruction output, resulting in $|A_L|$ instruction input-output pairs. 
We denote the sets of the instruction data from ID, title, and attribute facets as $\mathcal{D}_{ID}$, $\mathcal{D}_{title}$, and $\mathcal{D}_{attr}$, respectively.  
Moreover, to explicitly distinguish different facet data, we add a facet prefix after the instruction input as shown in Figure~\ref{fig:reconstruced_data}.

\vspace{2pt}
Based on the reconstructed instruction data 
$\mathcal{D}_{instruct} =\mathcal{D}_{ID}\cup \mathcal{D}_{title}\cup\mathcal{D}_{attr}$, 
the LLM is optimized via Eq. (\ref{eq:pre_finetuning_objective}). Note that we only employ a single LLM in TransRec for the instruction tuning.

\subsection{Multi-facet Generation Grounding}
After instruction tuning, the next step of TransRec is generation grounding (see Figure~\ref{fig:method_generation}), which aims to deliver in-corpus item recommendations based on the user's historical interactions. 
\subsubsection{\textbf{Position-free Constrained Generation}}
Out-of-corpus identifiers and over-reliance on the quality of initially generated tokens are two critical problems in the generation process. 
To tackle the issues, we consider LLMs to conduct position-free constrained generation. 
Remarkably, we introduce FM-index~\cite{ferragina2000opportunistic}, a specialized data structure, that simultaneously supports position-free and constrained generation. 

\begin{figure}[t]
\vspace{-0.5cm}
\setlength{\abovecaptionskip}{0.0cm}
\setlength{\belowcaptionskip}{-0.1cm}
\centering
\includegraphics[scale=0.42]{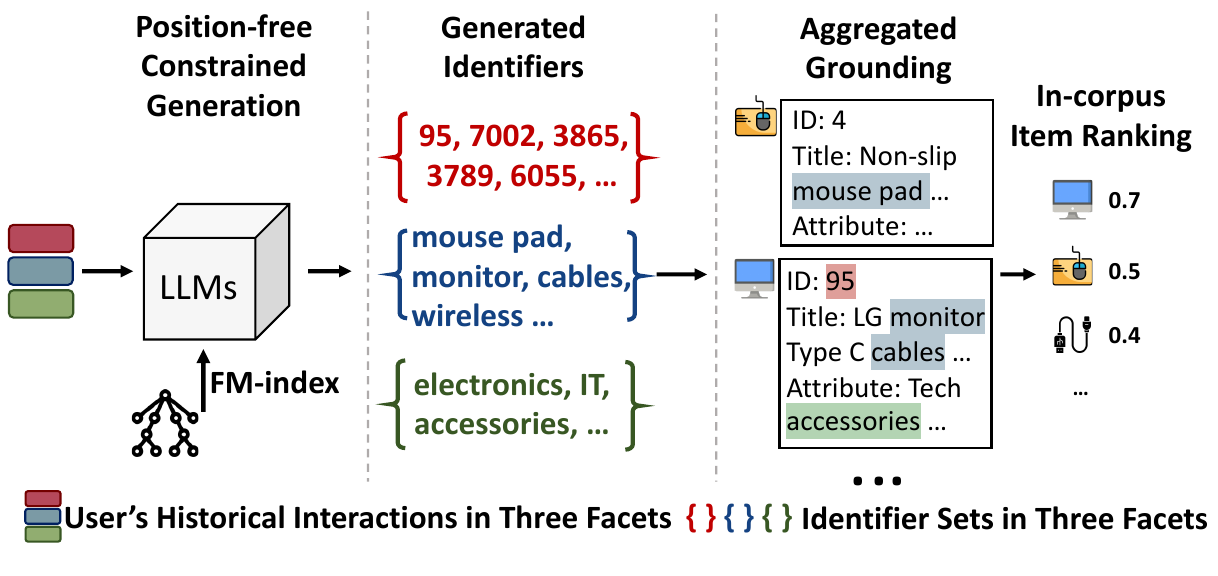}
\caption{{Demonstration of the generation grounding step in TransRec. Red, blue, and green denote the facets of ID, title, and attribute, respectively. }}
\label{fig:method_generation}
\end{figure}


\vspace{3pt}
\noindent$\bullet\ $\textbf{FM-index.} 
FM-index is a special prefix tree that supports search from any position~\cite{bevilacqua2022autoregressive}. 
This capability enables FM-index to 1) find all valid successor tokens of a given token; and 
2) allow the generation to start from any token of the valid identifiers\footnote{Other potential Trie algorithms in existing constrained generation merely allow starting from the first token of valid identifiers~\cite{chu2023leveraging,hao2022cgf}.}. 
Specifically, taking the item in Figure~\ref{fig:intro_ourmethod}(a) as an example, we flatten the multi-facet identifier as ``<IDS> 1023 <IDE> Urban Decay Eyeshadow Palette Naked Heat <AS> Makeup <AE> <AS> Eyes <AE>'', where ``<IDS>, <IDE>, <AS>, <AE>'' are the special tokens that indicate the start and the end of each ID and fine-grained attribute, respectively. The flattened identifier will then be stored in the Wavelet Tree~\cite{grossi2003high}.
Given a start token (\eg ``BOS'') or a token sequence, the FM-index can find a list of all possible successor tokens in $O(V\log(V))$, where $V$ is the vocabulary size of the LLMs (refer to Appendix~\ref{app:fm_index} for detailed explanations). 

\vspace{2pt}
\noindent$\bullet\ $\textbf{Identifier generation.} 
Given the user's historical interactions in the format of instruction input as in Figure~\ref{fig:reconstruced_data}, TransRec generates valid identifiers in each facet via constrained beam search~\cite{de2020autoregressive} based on the FM-index. 
Notably, the special tokens are utilized to indicate which facet is being generated. 
By constraining the starting token of generation, \eg ``<IDS>'', and the ending token of generation, \eg ``<IDE>'', TransRec generates a set of valid ID identifiers and attribute identifiers that belongs to the items (see Figure~\ref{fig:method_generation}). 
Besides, TransRec generates valid substrings of title identifiers from any position through FM-index, and thus we do not need to set special start and end tokens for the title. 
The position-free generation ability of LLMs is also enhanced by the instruction tuning process, where substrings are set as the instruction output (\cf Section~\ref{sec:multi-facet-identifier}). 
We follow~\cite{bevilacqua2022autoregressive} to keep track of all the partially decoded sequences and obtain a set of generated identifiers for each facet. 
We denote the generated identifiers for ID, title, and attribute facets as $\mathcal{P}_g$, $\mathcal{T}_g$, and $\mathcal{A}_g$, respectively.

\subsubsection{\textbf{Aggregated Grounding}}
To ground the generated identifiers to the in-corpus items and also rank in-corpus items for recommendations, we introduce an aggregated grounding module, which contains intra-facet and inter-facet aggregations. 

\vspace{2pt}
\noindent$\bullet\ $\textbf{Intra-facet aggregation.} 
We first ground the generated identifiers within each facet to the in-corpus items. 
Specifically, given the generated identifiers of one facet (\eg $\mathcal{T}_g$), we can aggregate them to the in-corpus item based on their coverage on each item's identifier (\eg $\mathcal{T}_g\cap T$). 
However, directly summing up the identifier scores\footnote{Here, the identifier score is the probability of generated identifier given by LLMs.} in the coverage set is infeasible due to the monotonic probability decrease in autoregressive generation~\cite{bevilacqua2022autoregressive}. 
For example, ``51770'' will have a smaller probability than ``517'', thus hindering the accurate grounding. 
To address this issue, we follow~\cite{bevilacqua2022autoregressive} to balance the scores by integrating token frequency: 
\begin{equation}\small\label{eq:grounding_length_weighted}
    s(\hat{y}) = \max\{0, \log\frac{P(\hat{y}|x)(1-P(\hat{y}))}{P(\hat{y})(1-P(\hat{y}|x))}\},
\end{equation}
where $\hat{y}$ is the generated identifier, $P(\hat{y}|x)$ is the token score given by the LLMs, and $P(\hat{y})$ is the unconditional probability that measures the frequency of tokens. 
For intuitive understanding, Eq. (\ref{eq:grounding_length_weighted}) can mitigate the issue by upweighting the tokens that are more distinctive, \ie less frequent. 
We obtain the unconditional probability $P(\hat{y})$ by
\begin{equation}\small\label{eq:norm_frequency}
    P(\hat{y})=\frac{F(\hat{y},\Tilde{\mathcal{I}})}{\sum_{d\in\Tilde{\mathcal{I}}}|d|},
\end{equation}
where $F(\hat{y},\Tilde{\mathcal{I}})$ represents the number of token $\hat{y}$ in the identifier corpus $\Tilde{\mathcal{I}}$. 
Then, the grounding score for item $i$ corresponding to user $u$ in facet $f$ is
\begin{equation}\small\label{eq:intra_facet_score}
    s_f(u,i) = \sum_{\hat{y}\in \Tilde{i}}(s(\hat{y}))^\gamma, 
\end{equation}
where $\Tilde{i}$ is the multi-facet identifiers of the item, and $\gamma$ is a hyper-parameter to control the strength of intra-facet grounding scores. 

\vspace{2pt}
\noindent$\bullet\ $\textbf{Inter-facet aggregation.} 
To aggregate the grounding scores from three facets, we should consider the disparate influence of each facet in different scenarios. 
For instance, when seeking books, the titles may become crucial, while for restaurants, the category (\eg Mexican cuisine) possibly takes precedence.
As such, TransRec balances the strength of each facet for the final ranking. 
Formally, the final grounding score for user $u$ and item $i$ is obtained by: 
\begin{equation}\small\label{eq:inter_facet_score}
    s(u,i) = \sum_{f}s_f(u,i)+b_{f},
\end{equation}
where $f\in\{ID, title, attribute\}$, and $b_f$ is the bias hyper-parameter that balances the strength between facets. According to the final grounding scores, we can obtain a ranking list of in-corpus items as in Figure~\ref{fig:method_generation} and return top-ranked items as recommendations. 

\vspace{3pt}
\noindent$\bullet\ $\textbf{Instantiation.} TransRec is a model-agnostic method that can be applied to any backbone LLMs and diverse tuning techniques. To investigate the feasibility of TransRec on LLMs with different sizes and architectures, we instantiate TransRec on two LLMs, \ie BART-large~\cite{lewis2019bart} and LLaMA-7B~\cite{touvron2023llama}. BART-large is an encoder-decoder model with 406M parameters, and LLaMA-7B is a decoder-only model with 7B parameters. 


%% file: 3_exp.tex
\section{Experiments}\label{sec:experiments}
In this section, we conduct experiments on three real-world datasets to answer the following research questions:
\begin{itemize}[leftmargin=*]
    \item \textbf{RQ1:} How does our proposed TransRec perform compared to both traditional and LLM-based recommenders?
    \item \textbf{RQ2:} How does TransRec perform under few-shot setting, on both warm- and cold-start recommendation? 
    \item \textbf{RQ3:} How does each component of TransRec (\eg each facet of identifier, constrained generation, and aggregated grounding) affect its effectiveness?

\end{itemize}
\subsection{Experimental Settings}

\subsubsection{\textbf{Datasets}}
We conduct experiments on three popular benchmark datasets: 
1) \textbf{Beauty} is the collection of user interactions with beauty products from Amazon review datasets\footnote{\url{ https://jmcauley.ucsd.edu/data/amazon/.}}. 
2) \textbf{Toys} is also one representative recommendation dataset drawn from Amazon review datasets, where each toy product has substantial meta information. 
3) \textbf{Yelp}\footnote{\url{https://www.yelp.com/dataset.}} is a popular restaurant dataset with rich user interactions on extensive dining places. 
For the three datasets, we assign each item a random unique numeric ID. We then use the assigned IDs, item titles, and item categories for the ID, title, and attribute facets for TransRec, respectively. 
Following~\cite{geng2022recommendation}, we adopt the leave-one-out strategy\footnote{For each user's interactions, we use the last item as the testing item, the item before the last item as the validation item, and the rest as training items.} 
to split the datasets into training, validation, and testing sets. In addition, we consider two training settings: 
1) \textbf{full training} uses all users' interactions in the training set to train the models; and 
2) \textbf{few-shot training} randomly selects $N$ users' interactions to train the models, where $N=1024$ or $2048$. 
The statistics of the datasets are summarized in Table~\ref{tab:datasets} in Appendix.

\subsubsection{\textbf{Baselines}} 
We compare TransRec with both traditional recommenders (MF, LightGCN, SASRec, ACVAE, and DCRec) and LLM-based recommenders (P5, SID, SemID+IID, CID+IID, and TIGER). 
1) \textbf{MF}~\cite{rendle2009bpr} is one of the most representative collaborative filtering models, which decomposes the user-item interactions into the user and the item matrices. 2) \textbf{LightGCN}~\cite{he2020lightgcn} is a graph-based model which linearly propagates the user and item representations from the neighborhood. 3) \textbf{SASRec}~\cite{kang2018self} is a representative sequential recommender model that adopts self-attention mechanism to learn the item dependency from user's interactions. 4) \textbf{ACVAE}~\cite{xie2021adversarial} incorporates contrastive learning and adversarial training into the VAE-based sequential recommender model. 
5) \textbf{DCRec}~\cite{dcrec2023} employs 
contrastive learning and disentangles user interest and conformity to mitigate the popularity bias. 6) \textbf{P5}~\cite{geng2022recommendation} is a unified framework, which uses additional data (\eg user's reviews) for pre-training of LLMs on various tasks. {We train P5 on sequential tasks for a fair comparison and assign random numeric IDs to items to prevent potential data leakage issue (\cf Appendix~\ref{app:data_leakage}).} 7) \textbf{SID}~\cite{hua2023index} leverages collaborative information by sequential indexing. The items interacted consecutively by a user are assigned consecutive numeric IDs. 8) \textbf{SemID+IID}~\cite{hua2023index} designs the numeric ID based on the items' meta information such as attributes, where items with similar semantics have similar numeric IDs. 9) \textbf{CID+IID}~\cite{hua2023index} considers the co-occurrence matrix of items to design the numeric ID, where items that co-occur in user-item interactions will have similar numeric IDs. 10) \textbf{TIGER}~\cite{rajput2023recommender} generates item identifier through a trainable codebook, which utilizes the item title and item descriptions to create new item tokens entailed by semantics.

\begin{table*}[t]
\vspace{-0.2cm}
\setlength{\abovecaptionskip}{0.05cm}
\setlength{\belowcaptionskip}{0cm}
\caption{Overall performance comparison between the baselines and TransRec instantiated on BART on three datasets. The best results are highlighted in bold and the second-best results are underlined. * implies the improvements over the second-best results are statistically significant ($p$-value < 0.01) under one-sample t-tests.}
\setlength{\tabcolsep}{2.7mm}{
\resizebox{\textwidth}{!}{
\begin{tabular}{l|cccc|cccc|cccc}
\toprule
 & \multicolumn{4}{c|}{\textbf{Beauty}} & \multicolumn{4}{c|}{\textbf{Toys}} & \multicolumn{4}{c}{\textbf{Yelp}} \\                 
 \textbf{Model} & \textbf{R@5} & \textbf{R@10} & \textbf{N@5} & \textbf{N@10} & \textbf{R@5} & \textbf{R@10} & \textbf{N@5} & \textbf{N@10} & \textbf{R@5} & \textbf{R@10} & \textbf{N@5} & \textbf{N@10} \\ \hline\hline
\textbf{MF} & 0.0294 & 0.0474 & 0.0145 & 0.0191 & 0.0236 & 0.0355 & 0.0153 & 0.0192 & 0.0220 & 0.0381 & 0.0138 & 0.0190 \\
\textbf{LightGCN} & 0.0305 & 0.0511 & 0.0194 & 0.0260 & 0.0322 & 0.0508 & 0.0215 & 0.0275 & 0.0255 & {\ul 0.0427} & 0.0163 & 0.0218 \\
\textbf{SASRec} & 0.0380 & 0.0588 & 0.0246 & 0.0313 & 0.0470 & 0.0659 & 0.0312 & 0.0373 & 0.0183 & 0.0296 & 0.0116 & 0.0152 \\
\textbf{{DCRec}} & 0.0452 & 0.0635 & 0.0327 & 0.0385 & {\ul 0.0498} & 0.0674 & 0.0335 & 0.0406 & 0.0207 & 0.0328 & 0.0115 & 0.0154 \\ 
\textbf{ACVAE} & {\ul 0.0503} & {\ul 0.0710} & {\ul 0.0356} & {\ul 0.0422} & 0.0488 & {\ul 0.0679} & {\ul 0.0350} & {\ul 0.0411} & 0.0211 & 0.0356 & 0.0127 & 0.0174 \\ \midrule
\textbf{P5} & 0.0059 & 0.0107 & {0.0033} & {0.0048} & {0.0031} & {0.0069} & {0.0022} & {0.0034} & {0.0039} & {0.0062} & {0.0024} & {0.0031} \\
\textbf{SID} & 0.0350 & 0.0494 & 0.0254 & 0.0301 & 0.0164 & 0.0218 & 0.0120 & 0.0139 & 0.0218 & 0.0332 & 0.0161 & 0.0187 \\ 
\textbf{SemID+IID} & 0.0290 & 0.0429 & 0.0200 & 0.0245 & 0.0145 & 0.0260 & 0.0069 & 0.0123 & 0.0196 & 0.0304 & 0.0141 & 0.0160 \\ 
\textbf{CID+IID} & {0.0484} & {0.0703} & 0.0337 & 0.0412 & 0.0169 & 0.0276 & 0.0104 & 0.0154 & {\ul 0.0265} & 0.0417 & {\ul 0.0184} & {\ul 0.0233} \\
\textbf{TIGER} & 0.0377 & 0.0567 & 0.0249 & 0.0310 & 0.0278 & 0.0426 & 0.0176 & 0.0223 & 0.0183 & 0.0298 & 0.0119 & 0.0156 \\ 
\cellcolor{gray!16}\textbf{TransRec-B} & \cellcolor{gray!16}\textbf{0.0504} & \cellcolor{gray!16}\textbf{0.0735*} & \cellcolor{gray!16}\textbf{0.0365*} & \cellcolor{gray!16}\textbf{0.0450*} & \cellcolor{gray!16}\textbf{0.0518*} & \cellcolor{gray!16}\textbf{0.0764*} & \cellcolor{gray!16}\textbf{0.0360*} & \cellcolor{gray!16}\textbf{0.0420*}
& \cellcolor{gray!16}\textbf{0.0354*} & \cellcolor{gray!16}\textbf{0.0457*} & \cellcolor{gray!16}\textbf{0.0262*} & \cellcolor{gray!16}\textbf{0.0306*} \\ \bottomrule
\end{tabular}
}}
\label{tab:overall_performance}
\end{table*}

\vspace{2pt}
\noindent$\bullet\quad$\textbf{Evaluation.} 
We adopt the widely used metrics Recall@$K$ and NDCG@$K$ to evaluate the models~\cite{yang2023generic,liu2023joint}, where $K$ is $5$ or $10$. 

\subsubsection{\textbf{Implementation Details.}} 
We employ BART and LLaMA as backbone LLMs for TransRec, and we denote the two variants as ``TransRec-B'' and ``TransRec-L'', respectively. 
For TransRec-B, we follow~\cite{hua2023index,geng2022recommendation} to sample subsequences of user's interactions for training, which is widely used in sequential recommender models~\cite{hidasi2015session}. 
As for LLaMA, the training on subsequences is involved in the training objectives of decoder-only architecture~\cite{zhao2023survey}, 
and only uses the entire user sequence for training. 
Besides, for each user's interaction sequence, we iteratively discard the first item in the sequence until the length of instruction input does not exceed the maximum input length of LLMs (1024 for BART and 512 for LLaMA). 
TransRec is trained with Adam~\cite{kingma2014adam} (TransRec-B) and AdamW~\cite{loshchilov2017decoupled} 
(TransRec-L) on four NVIDIA RTX A5000 GPUs. 
We fully tune the model parameters of TransRec-B and perform the parameter-efficient fine-tuning technique LoRA~\cite{hu2021lora} to tune TransRec-L.  
For a fair comparison, we set the beam size to 20 for TransRec and all LLM-based baselines. 
Detailed hyper-parameter settings for baselines and TransRec are presented in Appendix~\ref{app:baseline_implementation}.

\subsection{Overall Performance (RQ1)}\label{sec:overall_performance}
The results of the baselines and TransRec with BART as the backbone model under the full training setting are presented in Table~\ref{tab:overall_performance}, from which we have the following observations: 
\begin{itemize}[leftmargin=*] 
    \item Among traditional recommenders, sequential methods (SASRec, ACVAE, DCRec) surpass non-sequential methods (MF and LightGCN) on both Beauty and Toys. 
    The better performance stems from the sequential modeling of the user's interaction sequence, which captures dynamic shifts in user interests and intricate item dependencies. 
    Moreover, ACVAE usually outperforms other traditional recommenders. This is because adversarial training and contrastive learning encourage high-quality user representation and enhance the discriminability between items. 
    
    \item CID+IID consistently yields better performance than SemID+IID, which is consistent with the findings in~\cite{hua2023index}. 
    This is reasonable since CID+IID leverages the co-occurrence of items to construct hierarchical numeric IDs, \ie items with similar interactions have similar IDs. As such, the IDs are integrated with collaborative information, which strengthens the key advantage of ID-based identifiers. 
    In contrast, SemID+IID simply constructs IDs based on items' meta information, \ie items with similar semantics have similar identifiers. However, this can lead to misalignment between item identifiers and user behavior, thus degrading the performance (\cf Section~\ref{sec:introduction}). 
    
    \item TIGER usually achieves comparable performance to most of the traditional recommenders and surpasses SemID+IID on both Beauty and Toys. 
    The better performance is attributed to 1) the additional utilization of description for capturing semantics; 
    and 2) the learnable codebook to learn nuanced semantics compared to the manually defined semantic IDs (SemID+IID). 
    Besides, P5 yields unsatisfactory performance on the three datasets. 
    We believe that the inconsistency with the observations in~\cite{geng2022recommendation} is due to the sequential indexing strategy, which potentially leads to data leakage~\cite{rajput2023recommender}. 
    
    \item TransRec consistently yields the best performance across the three datasets, validating the superiority of our proposed transition paradigm. 
    Notably, TransRec outperforms LLM-based recommenders by a large margin without requiring information on cold-start items to construct item identifiers. 
    This further demonstrates the strong generalization ability of TransRec (see more analysis on the generalization of TransRec in Appendix~\ref{app:group_evaluation}). 
    The superiority of TransRec is attributed to 
    1) the utilization of multi-facet identifiers, which simultaneously satisfies semantics and distinctiveness to leverage the rich knowledge in LLMs and capture salient item features; and 
    2) the constrained and position-free generation that guarantees in-corpus item generation and mitigates the over-reliance on initial tokens. 

\end{itemize}

\vspace{-2pt}
\subsection{In-depth Analysis} 

\begin{table}[t]
\setlength{\abovecaptionskip}{0.05cm}
\setlength{\belowcaptionskip}{-0.20cm}
\caption{Performance comparison under the few-shot setting. The bold results highlight the superior performance compared to the best LLM-based recommender baseline. ``TransRec-B'' and ``TransRec-L'' denote using BART and LLaMA as backbone LLM, respectively.}
\setlength{\tabcolsep}{3.2mm}{
\resizebox{0.48\textwidth}{!}{
\begin{tabular}{c|l|cccc}
\toprule
\multicolumn{1}{l|}{} &  & \multicolumn{2}{c}{\textbf{Warm}} & \multicolumn{2}{c}{\textbf{Cold}} \\
\multicolumn{1}{l|}{\textbf{N-shot}} & \textbf{Model} & \textbf{R@5} & \textbf{N@5} & \textbf{R@5} & \textbf{N@5} \\ \hline
\multirow{5}{*}{\textbf{1024}} & \textbf{LightGCN} & 0.0205 & 0.0125 & 0.0005 & 0.0003 \\
 & \textbf{ACVAE} & 0.0098 & 0.0057 & 0.0047 & 0.0026 \\
 & \textbf{CID+IID} & 0.0100 & 0.0066 & 0.0085 & 0.0071 \\
 & \textbf{TransRec-B} & 0.0042 & 0.0028 & 0.0029 & 0.0021 \\
 & \cellcolor{gray!16}\textbf{TransRec-L} & \cellcolor{gray!16}\textbf{0.0141} & \cellcolor{gray!16}\textbf{0.0070} & \cellcolor{gray!16}\textbf{0.0159} & \cellcolor{gray!16}\textbf{0.0097} \\ \midrule
\multirow{5}{*}{\textbf{2048}} & \textbf{LightGCN} & 0.0186 & 0.0117 & 0.0005 & 0.0004 \\
 & \textbf{ACVAE} & 0.0229 & 0.0136 & 0.0074 & 0.0044 \\
 & \textbf{CID+IID} & 0.0150 & 0.0101 & 0.0078 & 0.0062 \\
  & \textbf{TransRec-B} & 0.0057 & 0.0031 & 0.0045 & 0.0026 \\
 & \cellcolor{gray!16}\textbf{TransRec-L} & \cellcolor{gray!16}\textbf{0.0194} & \cellcolor{gray!16}\textbf{0.0112} & \cellcolor{gray!16}\textbf{0.0198} & \cellcolor{gray!16}\textbf{0.0124} \\ \bottomrule

\end{tabular}
}}
\vspace{-0.5cm}
\label{tab:few-shot}
\end{table}

\subsubsection{\textbf{Few-shot Training (RQ2)}}
To study how TransRec performs under limited data, we conduct few-shot training with randomly selected $N$ users' interactions, where $N$ is set as 1024 or 2048. 
To evaluate the models, 
we involve $N$ users in few-shot training (\ie warm-start users) and another randomly selected $N$ users that have not been seen in few-shot training (\ie cold-start users) for testing. 
In addition, we split the testing set into warm and cold sets, where interactions between warm-start users and warm-start items belong to the warm set, otherwise the cold set. 

We compare both TransRec-B and TransRec-L with competitive baselines. 
The results on warm and cold sets over Beauty\footnote{Results on Yelp and Toys with similar observations are omitted to save space.} are presented in Table~\ref{tab:few-shot}. 
It is observed that 
1) traditional methods yield competitive performance on warm-start recommendation. This is because these ID-based methods are effective in capturing collaborative information from user-item interactions. 
ACVAE yields better performance on cold sets, as it discards user embedding, enabling effective generalization for cold-start users. 2) Medium-sized LLMs (CID+IID and TransRec-B) struggle with few-shot training. This indicates that a considerable amount of data is necessary to adapt medium-sized LLMs to perform well on recommendation tasks, which is also consistent with previous work~\cite{bao2023bi}. 
The better performance of CID+IID compared to TransRec-B possibly arises from the utilization of all item information to assign identifiers, as opposed to our strict reliance solely on warm items. 
3) Notably, TransRec-L outperforms all the baselines, particularly surpassing by a large margin on the cold set. 
This highlights the remarkable generalization ability of recently emerged LLMs with vast knowledge base, enabling more effective adaptation to the recommendation task with limited data.
 

\subsubsection{\textbf{Ablation Study (RQ3)}}\label{sec:ablation}

\begin{figure}[t]
\vspace{-0.2cm}
\setlength{\abovecaptionskip}{0cm}
\setlength{\belowcaptionskip}{0cm}
\centering
\includegraphics[scale=0.42]{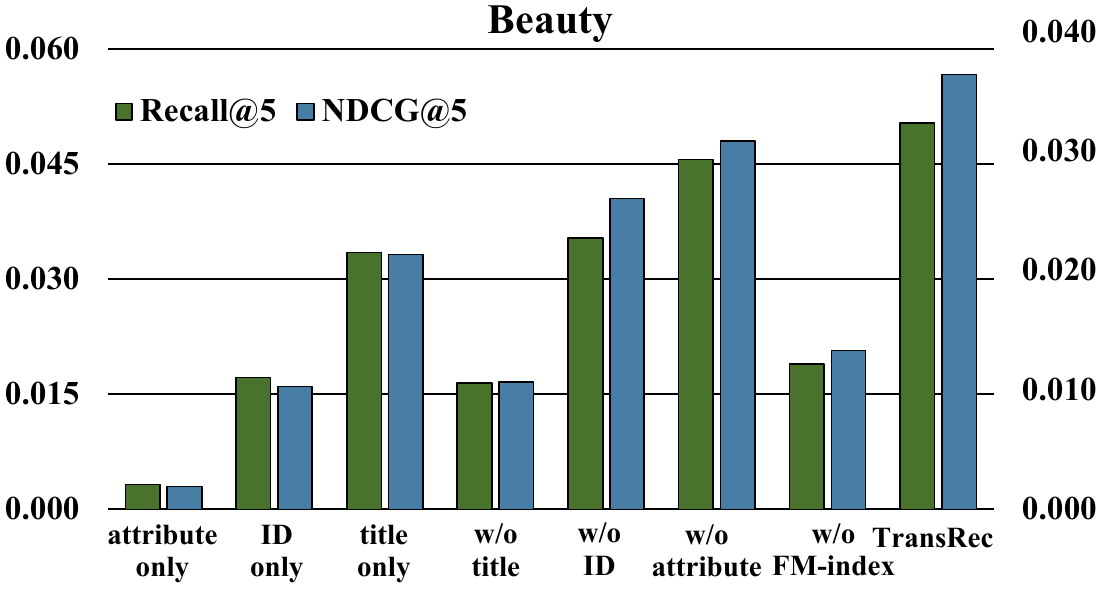}
\caption{Ablation study of each facet (\ie ID, title, and attribute) of multi-facet identifier and the FM-index.}
\label{fig:ablation}
\end{figure}

To analyze the effect of each facet in multi-facet identifiers, we remove the ID, title, and attribute separately, referred to as ``w/o ID'', ``w/o title'', and ``w/o attribute'', respectively. 
We also test TransRec with a single facet, \ie ID, title, and attribute, respectively. 
In addition, we disable the FM-index and conduct unconstrained generation, denoted as ``w/o FM-index''.  
Results on Beauty are presented in Figure~\ref{fig:ablation}. 
From the figure, we can find that: 
1) removing either ID, title, or attribute facet will decrease the performance, indicating the effectiveness of each facet in representing an item for LLM-based recommendation. 
2) Discarding the ID or title facet typically results in more significant performance reductions compared to removing the attribute facet. 
This is reasonable since removing IDs falls short of meeting distinctiveness criteria for identifiers, hindering LLMs from capturing salient features of items. 
Meanwhile, titles tend to possess more intricate semantics than attributes, exerting a more significant effect on enhancing recommendations. 
The crucial role of title facet is also indicated by the performance of TransRec with each single facet. 
4) It is not surprising that removing FM-index fails to give appropriate recommendations (inferior performance of ``w/o FM-index''), because it may generate our-of-corpus identifiers. This implies the necessity of position-free constrained generation. 

\vspace{2pt}
\noindent$\bullet\quad$\textbf{Effect of grounding strategies.}
We also compare the aggregated grounding module of TransRec with three potential grounding strategies. Following~\cite{bao2023bi}, we utilize LLMs to extract the representations of generated identifiers and the identifiers of in-corpus items, respectively. 
We then calculate the dot product, the negative L2 distance, and the cosine similarity between the generated identifiers and each in-corpus item as the grounding score for each in-corpus item, respectively, as three strategies. 
From the results in Table~\ref{tab:different_grounding}, we can observe that 
%
1) potential grounding strategies fail to yield satisfying results. 
This is because these strategies utilize the representations extracted from LLMs, thus relying heavily on the semantics similarity. As such, they may inaccurately ground the generated tokens that lack meaningful semantics to the valid identifiers. 
2) TransRec is more time-efficient compared to other grounding strategies. Because the three strategies introduce extra LLMs' forward process for extracting the representations, causing high computation burdens. 

\begin{table}[t]
\setlength{\abovecaptionskip}{0cm}
\setlength{\belowcaptionskip}{-0.30cm}
\caption{Performance comparison of different grounding strategies. ``Time'' denotes the time cost of the grounding process per 5,000 users.}
\setlength{\tabcolsep}{3mm}{
\resizebox{0.48\textwidth}{!}{
\begin{tabular}{lccccc}
\toprule
\multicolumn{6}{c}{\textbf{Beauty}} \\ 
\multicolumn{1}{l}{} & \textbf{R@5}$\uparrow$ & \textbf{R@10}$\uparrow$ & \textbf{N@5}$\uparrow$ & \textbf{N@10}$\uparrow$ & \textbf{Time}$\downarrow$ \\ \hline
\multicolumn{1}{l}{\textbf{Dot Product}} & 0.0016 & 0.0020 & 0.0013 & 0.0014 & 571.07 s \\
\multicolumn{1}{l}{\textbf{Cosine Sim}} & 0.0096 & 0.0121 & 0.0095 & 0.0101 & 578.89 s \\
\multicolumn{1}{l}{\textbf{L2 Distance}} & 0.0201 & 0.0212 & 0.0148 & 0.0161 & 577.64 s \\
\multicolumn{1}{l}{\textbf{TransRec}} & \textbf{0.0504} & \textbf{0.0735} & \textbf{0.0365} & \textbf{0.0450} & \textbf{218.43 s} \\ \bottomrule
\end{tabular}
}}
\label{tab:different_grounding}
\end{table}

\subsubsection{\textbf{Hyper-parameter Analysis}} 
\begin{figure}[t]
\vspace{-0.3cm}
\setlength{\abovecaptionskip}{-0.2cm}
\setlength{\belowcaptionskip}{-0.5cm}
  \centering 
  \subfigure{
    \includegraphics[height=1.4in]{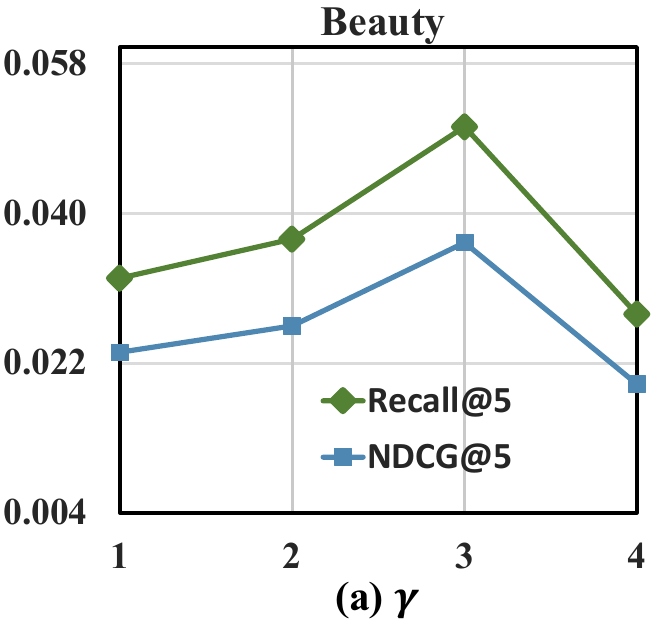}} 
  \subfigure{
    \includegraphics[height=1.4in]{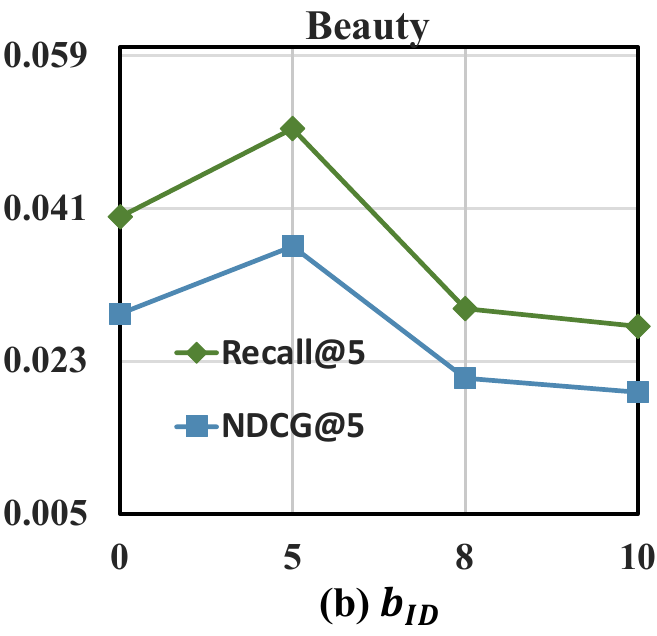}} 
  \caption{Effect of $\gamma$ and $b_{ID}$ in TransRec.}
  \label{fig:hp_analysis}
  
\end{figure}

We further study the sensitivity of hyper-parameters to facilitate future applications of TransRec. 
The performance of TransRec with different values of $\gamma$ and $b_{ID}$ are presented in Figure~\ref{fig:hp_analysis}. 
It is observed that 
1) as $\gamma$ varies from 1 to 3, the performance gradually improves. The possible reason is that the scales of original probabilities of LLMs are limited ($\gamma=1$), leading to insufficient discrepancies between identifiers and 
consequently reducing the informativeness of the ranking. 
2) However, it is crucial not to indiscriminately increase $\gamma$, 
as this may lean towards recommendations of representative items. 
3) We should carefully choose $b_{ID}$ to balance the strength between facets since a small $b_{ID}$ weakens the consideration of salient features while a large $b_{ID}$ might undermine other facets, thus hurting recommendations. 
Analysis of $b_{title}$ and $b_{attribute}$ are presented in Appendix~\ref{app:hp_analysis}.

%% file: 4_related_work.tex
\section{Related Work}\label{sec:related_work}
\subsection{LLMs for Recommendation}
Recently, leveraging LLMs for recommendation has received much attention due to LLMs' rich world knowledge, strong reasoning, and generalization abilities~\cite{wu2021empowering,zhao2022resetbert4rec,geng2022path,zhang2023prompt,li2023text,wang2022towards,li2022pear,borisov2022language}. 
Existing work on LLMs for recommendation can be mainly divided into two groups. 
1) LLM-enhanced recommenders~\cite{zou2021pre,hou2022towards,guo2023few,chen2023palr}, which consider LLMs as powerful feature extractors for enhancing the feature representation of users and items. 
Another line of research lies in 
2) LLM-based recommenders~\cite{peng2024towards,li2024survey,wang2024learnable}, which directly leverage the LLMs as recommender systems. 
In the wake of ChatGPT's release and its remarkable prowess in reasoning, early studies delve into LLMs' zero-shot/few-shot recommendation capabilities through in-context learning~\cite{dai2023uncovering,hou2023large,wang2023zero}. 
However, due to the intrinsic gap between the LLMs' pre-training and recommendation tasks~\cite{chen2023palr,lin2023can}, the recommendation abilities are somehow limited by merely using in-context learning~\cite{liu2023chatgpt}. 
Therefore, to narrow the gap and elicit the strong capabilities of LLMs, recent studies utilize instruction tuning for LLMs to improve the performance~\cite{zhang2023recommendation,bao2023tallrec,lin2024data}. 
In this work, we highlight the two fundamental steps of LLM-based recommenders to bridge the item and language and spaces, showing existing problems and the key objectives in the two steps. 

\subsection{Item Indexing and Generation Grounding}
To bridge the item space and the language space, the two vital steps are item indexing and generation grounding.  
Previous indexing methods can be categorized into two groups. 
1) ID-based identifiers represent each item by a unique numeric ID, 
to learn salient features from 
user-item interactions~\cite{li2023ctrl,hua2023index,yuan2021one}. 
2) Description-based identifiers employ item descriptions to represent an item~\cite{zhang2023recommendation,cui2022m6}. 
Nevertheless, ID-based identifiers lack semantics, while identifier-based identifiers lack adequate distinctiveness. 
To alleviate the issues,~\cite{hua2023index,rajput2023recommender} propose to design IDs based on descriptions, which however is essentially a description-based identifier and yields less satisfying recommendations (\cf Section~\ref{sec:overall_performance}). 
Although~\cite{yuan2023go} and~\cite{li2023exploring} compare the two types of identifiers independently in a discriminative manner, TransRec employs the generative approach to leverage both semantics and distinctiveness for effective item indexing. 
{A concurrent study~\cite{li2023multiview} also explores multiview identifiers in passage ranking. However, it solely considers the semantics to strengthen the correlation between query and passages. In contrast, our study takes the initial endeavor to investigate multi-facet identifiers in LLM-based recommendation, highlighting the distinctiveness as a criterion for identifier design to capture the crucial CF knowledge. }

The generation grounding step involves the generation of token sequences and the grounding to the in-corpus items. 
However, this step has received little scrutiny in previous studies. They mainly conduct autoregressive generation over the whole vocabulary and ground the generated tokens by exact matching, which may lead to out-of-corpus recommendations. 
\cite{cui2022m6} and~\cite{bao2023bi} mitigate the out-of-corpus issue by utilizing the representations obtained by LLMs to match between generated tokens and in-corpus items, \eg L2 distance. 
However, these approaches inevitably suffer from high computational burdens as they require extra computations for representation extraction. 
Different from previous work, we propose an effective generation grounding step, which utilizes a specialized data structure to achieve guaranteed in-corpus and enhanced recommendations. 

%% file: 5_conclusion.tex
\section{Conclusion and Future Work}\label{sec:conclusion}
In this work, we highlighted the two fundamental steps for LLM-based recommenders: item indexing and generation grounding. 
To make full utilization of LLMs and strengthen the generalization ability of LLM-based recommenders, we posited that item identifiers should pursue distinctiveness and semantics. 
In addition, we considered position-free constrained generation for LLMs to yield accurate recommendations. 
To pursue these objectives, we proposed a novel transition paradigm, namely TransRec, to seamlessly bridge the language space and the item space. 
We utilized multi-facet identifiers to represent an item from ID, title, and attribute facets simultaneously. 
Besides, we employed the FM-index to guarantee high-quality generated identifiers. 
Furthermore, we introduced an aggregated grounding module to ground the generated identifiers to the items. 
Empirical results on three real-world datasets under diverse settings
validated the superiority of TransRec in improving recommendation accuracy and generalization ability. 

This work highlights the key objectives of indexing approaches and generation grounding strategies, leaving many promising directions for future exploration. 
Particularly, 1) although incorporating ID, title, and attribute is effective, it is worthwhile to automatically construct multi-facet identifiers to reduce the noises in natural descriptions;  
and 2) it is meaningful to devise better strategies for grounding modules, to effectively combine the ranking scores from different facets, such as using neural models in a learnable manner. 

%% file: 6_SI.tex
\section{Appendix}
\label{sec:appendix}

\subsection{FM-index}\label{app:fm_index}
FM-index is a compressed suffix array. Given the original text, \eg item title, it is first transformed into a matrix based on the Burrows-Wheeler Transform (BWT)~\cite{schindler1997fast}, which turns the original text into a matrix sorted lexicographically. For example, given a string $XYXZ$, the transformed matrix is: 

\begin{table}[h]
\begin{tabular}{lllll}
\$ & X & Y & X & Z \\
X & Y & X & Z & \$ \\
X & Z & \$ & X & Y \\
Y & X & Z & \$ & X \\
Z & \$ & X & Y & X,
\end{tabular}
\caption{Example of transformed matrix based on BWT.}
\label{tab:fm_index}
\end{table}

As shown in the example, the first column is the repeated token sorted lexicographically, and the last column is called the string’s BWT. 
Based on the matrix, it can then access every token through self-indexing and can find all valid token successors. 
Notably, only the first and last columns are explicitly stored in the FM-index, contributing to its space efficiency. 
Then, the storage of these columns is managed by the Wavelet Tree, a hierarchical structure that employs wavelet transformations for compact encoding and rapid querying of the FM-index. 
This integrated approach enhances the overall performance and scalability of the data structure. 
More detailed explanations can also be found in~\cite{bevilacqua2022autoregressive}.

\subsection{Hyper-parameter Settings}\label{app:baseline_implementation}
The best hyper-parameters of models are selected by Recall on the validation set. For the traditional recommender models, we search the embedding size, learning rate, and weight decay from $\{16,32,64,128,256\}$, $\{1e^{-4},1e^{-3},1e^{-2}\}$, and $\{1e^{-4},1e^{-3},1e^{-2},1e^{-1}\}$, respectively. The search scopes for some model-specific hyper-parameters are as follows. For LightGCN, we tune the number of GCN layers and the dropout ratio in $\{1,2,3,4,5\}$ and $\{0.1,0.3,0.5\}$, respectively. We search the training sequence length in $\{10, 20, 50, 100\}$ for both SASRec and ACVAE. For SASRec, we select the number of attention blocks in $\{1,2,3\}$ and the number of attention heads in $\{1,2,4,8\}$. For ACVAE, the weight of contrastive loss term $\beta$ is searched in $\{0.1,0.3,0.5,0.7\}$. 
For LLM-based recommenders~\cite{hua2023index,rajput2023recommender}, we follow the searching scopes stated in their papers. 
For TransRec, we set the sampling number of substrings $K$ as $5$. The scaling factor for the intra-facet grounding scores $\gamma$ and the bias terms for the intra-facet aggregation $b_{ID}$, $b_{title}$, and $b_{attribute}$ are searched in ranges of $\{1,2,3,4\}$, $\{0,5,8,10\}$,$\{-2,0,2,5\}$, and $\{2,5,7,10\}$, respectively. 

\subsection{Performance of TransRec on T5-small}
To achieve fair comparisons, we employ a consistent backbone LLM and prompt template across all LLM-based methods. Specifically, we employ T5-small to instantiate SID, SemID+IID, CID+IID, and TransRec. Besides, to avoid the bonus from multiple prompt templates, we implement all methods with the same single prompt. 
From the results in Table~\ref{app:performance-t5}, 
we can find that TransRec outperforms baselines by a large margin, which demonstrates the effectiveness of multi-facet identifier and generation grounding. 

\begin{table}[h]
\setlength{\abovecaptionskip}{0.05cm}
\setlength{\belowcaptionskip}{0cm}
\caption{Performance comparison between TransRec and various indexing methods with the same backend model T5. }
\setlength{\tabcolsep}{2mm}{
\resizebox{0.48\textwidth}{!}{
\begin{tabular}{lcccc}
\hline
\multicolumn{5}{c}{\textbf{Beauty}} \\
 & \textbf{Recall@5} & \textbf{Recall@10} & \textbf{NDCG@5} & \textbf{NDCG@10} \\ \midrule
\textbf{SID} & 0.0113 & 0.0215 & 0.0069 & 0.0101 \\
\textbf{SemID+IID} & 0.0089 & 0.0192 & 0.0056 & 0.0090 \\
\textbf{CID+IID} & 0.0125 & 0.0230 & 0.0078 & 0.0112 \\
\textbf{TransRec-T5} & 0.0325 & 0.0493 & 0.0224 & 0.0278 \\ \hline
\end{tabular}
}}
\label{app:performance-t5}
\end{table}

\subsection{Cost Analysis of FM-index}
We analyze the storage and time costs of FM-index with results presented in Table~\ref{tab:fm-index-costs} and Table~\ref{tab:fm-index-update}, respectively. 
From the tables, we can find that both the storage costs, time costs, and update costs are trivial, facilitating the real-world application of FM-index. 

\begin{table}[h]
\setlength{\abovecaptionskip}{0.05cm}
\setlength{\belowcaptionskip}{0cm}
\caption{Storage and time costs of creating FM-index.}
\setlength{\tabcolsep}{2mm}{
\resizebox{0.4\textwidth}{!}{
\begin{tabular}{lcccccc}
\hline
\textbf{\# Item} & \textbf{10k} & \textbf{20k} & \textbf{30k} & \textbf{40k} & \textbf{50k} & \textbf{100k} \\ \hline
\textbf{Storage (MB)} & 1.2 & 1.8 & 3.8 & 5.0 & 6.2 & 13.0 \\
\textbf{Time costs (s)} & 6.2 & 11.9 & 18.3 & 23.6 & 28.7 & 55.2 \\ \hline
\end{tabular}
}}
\label{tab:fm-index-costs}
\end{table}

\begin{table}[h]
\setlength{\abovecaptionskip}{0.05cm}
\setlength{\belowcaptionskip}{0cm}
\caption{Time costs of adding 1k items to FM-index.}
\setlength{\tabcolsep}{2mm}{
\resizebox{0.4\textwidth}{!}{
\begin{tabular}{lcccccc}
\hline
\textbf{\# Existing item} & \textbf{10k} & \textbf{20k} & \textbf{30k} & \textbf{40k} & \textbf{50k} & \textbf{100k} \\
\textbf{Time costs (s)} & 1.1 & 1.1 & 1.6 & 1.6 & 2.0 & 3.8 \\ \hline
\end{tabular}
}}
\label{tab:fm-index-update}
\end{table}

\subsection{Constrained Generation with Trie}
We employ Trie, a data structure that supports generating valid tokens strictly from the first token of the item identifier during constrained generation. 
From the results in Table~\ref{tab:trie}, we can find that generating from any position outperforms generating from the first token, which is due to potential misalignment between the first token and the user preference. 
This validates the effectiveness of position-free generation supported by FM-index. 

\begin{table}[h]
\setlength{\abovecaptionskip}{0.05cm}
\setlength{\belowcaptionskip}{0cm}
\caption{Performance of constrained generation with Trie.}
\setlength{\tabcolsep}{2mm}{
\resizebox{0.4\textwidth}{!}{
\begin{tabular}{lcccc}
\hline
\multicolumn{5}{c}{\textbf{Beauty}} \\ \hline
 & \textbf{Recall@5} & \textbf{Recall@10} & \textbf{NDCG@5} & \textbf{NDCG@10} \\
\textbf{Trie} & 0.0186 & 0.0274 & 0.0118 & 0.0146 \\
\textbf{FM-index} & \textbf{0.0504} & \textbf{0.0735} & \textbf{0.0365} & \textbf{0.0450} \\ \hline
\end{tabular}
}}
\label{tab:trie}
\end{table}

\subsection{Potential Data Leakage Issue of P5}\label{app:data_leakage}
{As discussed in~\cite{rajput2023recommender}, the original sequential indexing method utilized in P5~\cite{geng2022recommendation} suffers from the potential data leakage issue. 
Specifically, P5 assigns consecutive numeric IDs to the interacted items in the user sequence, where the items in training and testing sets have a high probability of sharing the same token after tokenization. 
For instance, P5 may represent a user sequence as 
$[7391, 7392, \dots, 7398, 7399]$, where $7399$ is in the testing set. 
Based on the SentencePiece tokenizer~\cite{sennrich2016neural}, these numeric IDs will be tokenized into ``73'' and ``91'', ``73'' and ``92'', and so forth. 
As such, the item identifiers in the training and the testing sets will share the same token ``73''. 
This can lead to strong correlations between the historical interactions and next-interacted items, thus significantly benefiting the prediction accuracy of the testing item. 
However, such benefits are from the consecutive numbers, which are unattainable during the indexing process in real-world scenarios. 
}
{
To solve this issue, we adopt the datasets in P5 for experiments but rearrange the numeric ID for items with random numeric ID instead of consecutive IDs for both our method and P5. 
}

\subsection{User Group Evaluation}\label{app:group_evaluation}

\begin{figure}[h]
\vspace{-0.5cm}
\setlength{\abovecaptionskip}{-0.1cm}
\setlength{\belowcaptionskip}{-0.1cm}
  \centering 
  \hspace{-0.15in}
  \subfigure{
    \includegraphics[height=1.6in]{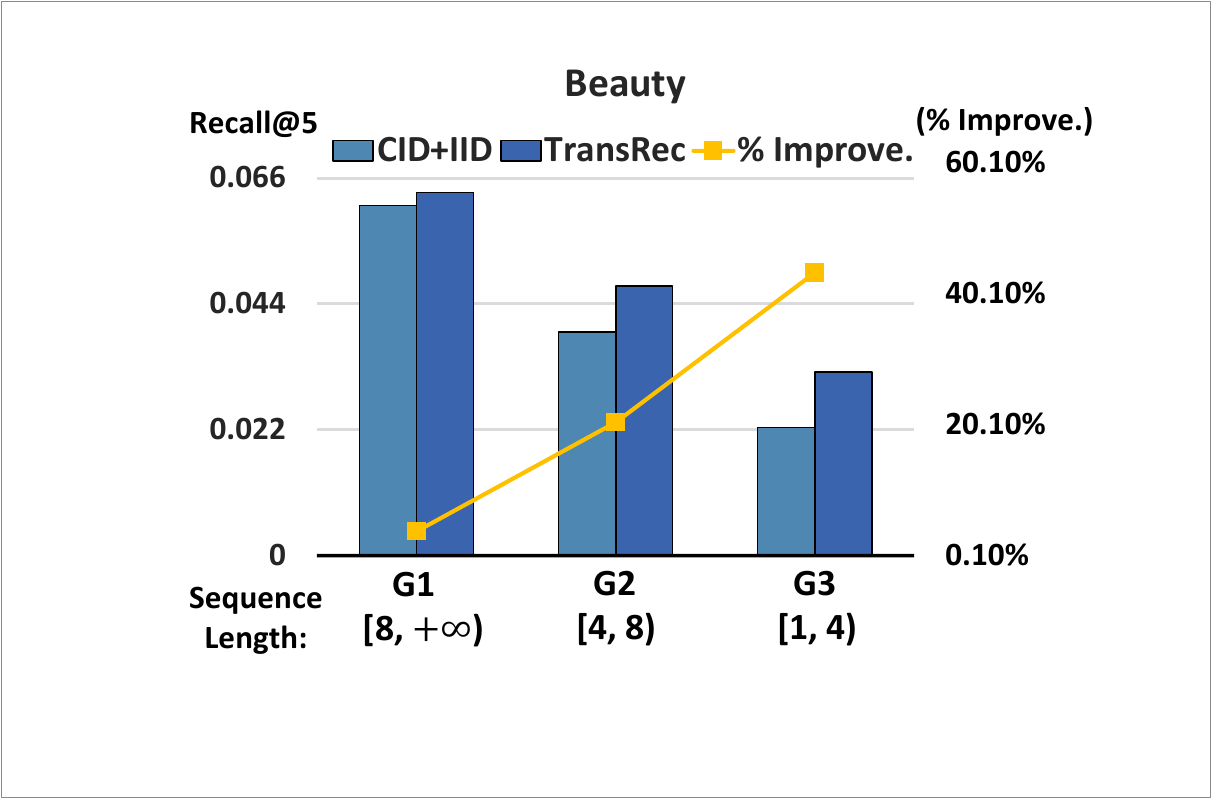}} 
  \caption{Performance over user groups with different lengths of historical interaction sequence.}
   \label{fig:group_evaluation}
\end{figure}

To analyze how TransRec improves the performance and the generalization ability of LLM-based recommenders, we test the performance of TransRec over different sparsity of users and compare it with the competitive baseline CID+IID. 
Specifically, we divide the users into three groups according to the sequence length of historical interactions. 
We select users with interactions larger or equal to 8 into group 1, denoted as ``G1''; and then split the rest of the users with interactions larger or equal to 4 into group 2, otherwise group 3, denoted as ``G2'', and ``G3'', respectively. 
As such, from G1 to G3, the user sparsity increases. 
The results of the three groups on Beauty are presented in Figure~\ref{fig:group_evaluation}. We can find that: 
1) From G1 to G3, the performance of both CID+IID and TransRec decreases. This makes sense because it can be difficult to capture the user preference shifts from only a small number of interactions. 
Nevertheless, 
2) TransRec consistently outperforms CID+IID under different levels of user sparsity. In particular, TransRec improves the performance of sparse users remarkably by a large margin (significant improvements on ``G3''), indicating the strong generalization ability of TransRec. 

\begin{figure}[h]
\vspace{-0.5cm}
  \centering 
  \hspace{-0.15in}
  \subfigure{
    \includegraphics[height=1.52in]{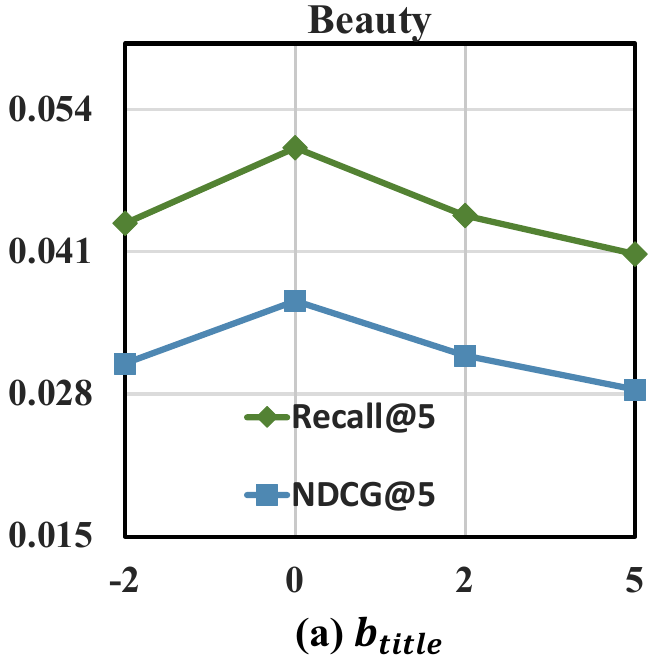}} 
  \subfigure{
    \includegraphics[height=1.52in]{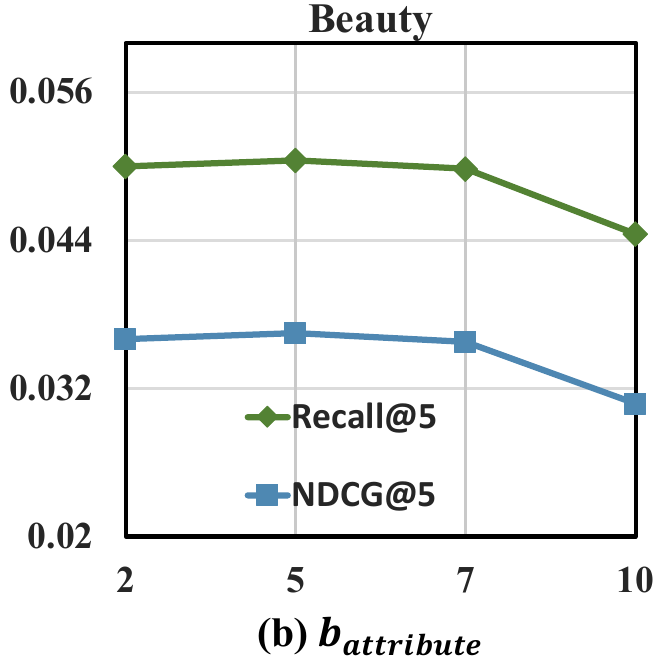}} 
\setlength{\abovecaptionskip}{-0.05cm}
\setlength{\belowcaptionskip}{-0.1cm}
\caption{Effect of $b_{title}$ and $b_{attribute}$ in TransRec.}
\label{fig:app_hp_analysis}
\end{figure}

\begin{table}[t]
\setlength{\abovecaptionskip}{0.05cm}
\setlength{\belowcaptionskip}{-0.05cm}
\caption{Statistics of three datasets.}
\setlength{\tabcolsep}{2mm}{
\resizebox{0.42\textwidth}{!}{
\begin{tabular}{lcccc}
\hline
\textbf{Dataset} & \multicolumn{1}{l}{\textbf{\# User}} & \multicolumn{1}{l}{\textbf{\# Item}} & \multicolumn{1}{l}{\textbf{\# Interaction}} & \multicolumn{1}{l}{\textbf{Density (\%)}} \\ \hline
\textbf{Beauty} & 22,363 & 12,101 & 198,502 & 0.0734 \\
\textbf{Toys} & 19,412 & 11,924 & 167,597 & 0.0724 \\
\textbf{Yelp} & 30,431 & 20,033 & 316,354 & 0.0519 \\ \hline
\end{tabular}
}}
\label{tab:datasets}
\end{table}

\subsection{Hyper-parameter Analysis}\label{app:hp_analysis} 

\noindent$\bullet\quad$\textbf{Effect of $b_{title}$. } 
We vary the bias for the title facet \ie $b_{title}$, and present the results in Figure~\ref{fig:app_hp_analysis}. 
It is observed that 
TransRec achieves the best performance when $b_{title}=0$, indicating that bias for the title facet may not necessarily need careful adjustment. One possible reason is that titles usually contain common words, resulting in a mild gap between the pre-training data and the titles. 
In contrast, IDs that are less common in pre-training data probably lead to a larger gap between the pre-training data and the identifiers, thereby requiring a larger bias to improve the strength of the ID facet. This is also evidenced by Figure~\ref{fig:hp_analysis}(b), where TransRec achieves the best performance when $b_{ID}=5$.   

\vspace{2pt}
\noindent$\bullet\quad$\textbf{Effect of $b_{attribute}$.} 
The results of different bias values for the attribute facet $b_{attribute}$ are reported in Figure~\ref{fig:app_hp_analysis}(b). 
From the results, we can find that 
1) gradually increasing the bias for the attribute facet does not affect the performance too much, indicating that TransRec might be less sensitive to the attribute strength. 
This is reasonable since attribute entails some coarse-grained semantics, such as category, and color, which can be shared by extensive items. 
2) Similar to the ID facet, strengthening either the attribute or title facet too much can hurt the performance. 
The reason is that letting the title or attribute facet dominate the grounding can decrease item discriminability. 